\documentclass[a4paper]{jpconf}
\usepackage{graphicx}
\usepackage{latexsym}
\usepackage{revsymb}
\usepackage{amsfonts}
\usepackage{amsmath}
\usepackage{amssymb}
\usepackage{mathtools}
\usepackage{bm}
\usepackage{ytableau}
\usepackage{multirow}
\usepackage{color}

\newcommand{\bra}[1]{\langle #1 |}
\newcommand{\ket}[1]{| #1 \rangle}
\newcommand{\braket}[2]{\langle #1 | #2\rangle}

\newcommand{\kz}{|\,0\,\rangle}
\newcommand{\ko}{|\,1\,\rangle}
\newcommand{\kp}{|+\rangle}
\newcommand{\km}{|-\rangle}
\newcommand{\kpm}{|\pm\rangle}
\newcommand{\kpi}{|\!+\!i\,\rangle}
\newcommand{\kmi}{|\!-\!i\,\rangle}
\newcommand{\kpmi}{|\!\pm\! i\,\rangle}

\newcommand{\0}{\underline{0}}
\newcommand{\1}{\underline{1}}
\newcommand{\2}{\underline{2}}

\ytableausetup{mathmode,boxsize=2.5ex,aligntableaux=center}
\newcommand{\G}{*(green)}
\newcommand{\W}{*(white)}
\begin{document}
%%%%%%%%%%%%%%%%%%%%%%%%%%%%%%%%%%%%%%%%%%%%%%%%%%%%%%%%%%%%%%%%%%%%%%%%%%%%%%%%%%%%%%%%%%%%%%%%%%%%
\title{Spekkens' Toy Model, Finite Field Quantum Mechanics, and the Role of Linearity}

\author{Lay Nam Chang, Djordje Minic, and Tatsu Takeuchi\footnote{Presenting Author}}

\address{Department of Physics, Virginia Tech, 850 West Campus Drive, Blacksburg 24061 USA}

\ead{laynam@vt.edu, dminic@vt.edu, takeuchi@vt.edu}

\begin{abstract}
We map Spekkens' toy model to a quantum mechanics defined over the finite field $\mathbb{F}_5$.
This allows us to define arbitrary linear combinations of the epistemic states in the model. 
For Spekkens' elementary system with only $2^2=4$ ontic states, the mapping is exact and the two
models agree completely.  However, for a pair of elementary systems 
there exist interesting differences between the entangled states of the two models.
\end{abstract}

%%%%%%%%%%%%%%%%%%%%%%%%%%%%%%%%%%%%%%%%%%%%%%%%%%%%%%%%%%%%%%%%%%%%%%%%%%%%%%%%%%%%%%%%%%%%%%%%%%%%
\section{Introduction}

The interpretation of Quantum Mechanics (QM) remains a subject of intense investigation and heated debate, 
as well as the source of puzzlement and intellectual discomfort for many.
To deepen our understanding of QM, 
various ontological models which posit some type of underlying physical ``reality'' to QM have been proposed. 
%An illuminating discussion on ontological models can be found in Ref.~\cite{Harrigan:2010}, where they are classified broadly into $\psi$-ontic, and $\psi$-epistemic models.
An illuminating discussion on ontological models of QM can be found in Harrigan and Spekkens \cite{Harrigan:2010}, 
in which the models are classified into two broad categories: $\psi$-ontic, and $\psi$-epistemic.
$\psi$-ontic refers to models in which different wave-functions $\psi$ correspond to different physical states of the system itself,
while $\psi$-epistemic refers to models in which different $\psi$'s correspond to different states of our (limited)
knowledge about the system. 
Both types of models have their pros and cons, and are often
unable to completely reproduce the predictions of canonical QM.
Nevertheless, scrutinizing where they succeed and where they fail 
will help us identify how or whether the models can be improved,
clarify the limitations of our ``classical'' thinking, and guide us to think more ``quantum-ly'' about QM.

In this note, we will look at a simple $\psi$-epsitemic ontological model proposed by Spekkens in Ref.~\cite{Spekkens:2007}.
The most elementary version of the model succeeds in reproducing the predictions of canonical QM of spin restricted to the six
states of $\ket{0}$ (spin up), $\ket{1}$ (spin down), $\ket{\pm}=\frac{1}{\sqrt{2}}(\ket{0}\pm\ket{1})$,
and $\ket{\!\pm\! i}=\frac{1}{\sqrt{2}}(\ket{0}\pm i\ket{1})$.
Furthermore, the two-``spin'' system in Spekkens' model possesses the analog of entanglement.
However, one aspect of QM which is missing from Spekkens' model is \textit{linearity}. 
Linearity, or the superposition principle, is what leads to the various mysteries of canonical QM, so
it is a property one would like to impart on any model of QM.
We argue that this can be done to Spekkens' model by mapping it to a QM defined
over the finite field $\mathbb{F}_5$ \cite{GQM:2013,GQM-Spin:2013,Chang:2013,Chang:2014,QFun:2014}.

%$\psi$-complete refers to models in which the wave-function $\psi$ describes the complete reality of the system, 
%and different wave-functions correspond to different realities.
%$\psi$-supplemented refers to hidden variable models in which different wave-functions correspond to different
%realities, but a set of hidden variables needs to be specified for a complete description.
%$\psi$-epistemic refers to models in which the wave-function $\psi$ describes our limit knowledge about reality,
%and different wave-functions 

%%%%%%%%%%%%%%%%%%%%%%%%%%%%%%%%%%%%%%%%%%%%%%%%%%%%%%%%%%%%%%%%%%%%%%%%%%%%%%%%%%%%%%%%%%%%%%%%%%
%%%%%%%%%%%%%%%%%%%%%%%%%%%%%%%%%%%%%%%%%%%%%%%%%%%%%%%%%%%%%%%%%%%%%%%%%%%%%%%%%%%%%%%%%%%%%%%%%%
\section{Spekkens' Toy Model}

We begin with a quick review of Spekkens' toy model \cite{Spekkens:2007}.

%%%%%%%%%%%%%%%%%%%%%%%%%%%%%%%%%%%%%%%%%%%%%%%%%%%%%%%%%%%%%%%%%%%%%%%%%%%%%%%%%%%%%%%%%%%%%%%%%%
\subsection{Ontic \& Epsitemic States}

Spekkens' \textit{knowledge balance principle} states that 
for a system which requires $2N$ bits of information to completely specify,
only $N$ bits of information may be known about the system at any time.
The state of the system is called the \textit{ontic state}, while the state of our
(limited) knowledge about the system is called the \textit{epistemic state}.

The most elementary system onto which Spekkens' knowledge balance principle could be applied
would be such that require $2$ bits of information to completely specify.
Such an elementary system would have $2^2 = 4$ ontic states which can be represented graphically as 
\begin{equation}
\begin{ytableau}
\W 1 & \W 2 & \W 3 & \W 4
\end{ytableau}
\end{equation}
The elementary system inhabits one and only one of these four ontic states at any time.
%When necessary, we use a black dot to indicate the ontic state the system is in:
%
%\begin{eqnarray}
%1 & \quad & \begin{ytableau} \W \bullet & \W & \W & \W \end{ytableau} \cr
%2 & \quad & \begin{ytableau} \W & \W \bullet & \W & \W \end{ytableau} \cr
%3 & \quad & \begin{ytableau} \W & \W & \W \bullet & \W \end{ytableau} \cr
%4 & \quad & \begin{ytableau} \W & \W & \W & \W \bullet \end{ytableau} 
%\end{eqnarray}

The epistemic states each represent 1 bit of knowledge about this system, which would only allow us to 
narrow down the possible ontic states to two:  $a\vee b$ where $a,b\in\{1,2,3,4\}$ and $a\neq b$.
$a\vee b$ and $b\vee a$ represent the same knowledge, so we will use the convention that $a<b$ for
all epistemic states $a\vee b$.\footnote{%
The symbol $\vee$ represents an OR.}
Therefore,
an elementary system possesses six epistemic states and these can be represented 
by shading the boxes where the actual ontic state may be:
\begin{eqnarray} 
1\vee 2 
& \quad & 
\begin{ytableau}
\G & \G & \W & \W 
\end{ytableau}
\;=\;
\begin{ytableau} \W \bullet & \W & \W & \W \end{ytableau}
\vee
\begin{ytableau} \W & \W \bullet & \W & \W \end{ytableau}
\cr
%%%
3\vee 4 
& \quad & 
\begin{ytableau}
\W & \W & \G & \G 
\end{ytableau}
\;=\;
\begin{ytableau} \W & \W & \W \bullet & \W \end{ytableau}
\vee
\begin{ytableau} \W & \W & \W & \W \bullet \end{ytableau}
\cr
%%%
1\vee 3 
& \quad & 
\begin{ytableau}
\G & \W & \G & \W  
\end{ytableau}
\;=\;
\begin{ytableau} \W \bullet & \W & \W & \W \end{ytableau}
\vee
\begin{ytableau} \W & \W & \W \bullet & \W \end{ytableau}
\cr
%%%
2\vee 4 
& \quad & 
\begin{ytableau}
\W & \G & \W & \G 
\end{ytableau}
\;=\;
\begin{ytableau} \W & \W \bullet & \W & \W \end{ytableau}
\vee
\begin{ytableau} \W & \W & \W & \W \bullet \end{ytableau}
\cr
%%%
2\vee 3 
& \quad & 
\begin{ytableau}
\W & \G & \G & \W 
\end{ytableau}
\;=\;
\begin{ytableau} \W & \W \bullet & \W & \W \end{ytableau}
\vee
\begin{ytableau} \W & \W & \W \bullet & \W \end{ytableau}
\cr
%%%
1\vee 4 
& \quad & 
\begin{ytableau}
\G & \W & \W & \G  
\end{ytableau}
\;=\;
\begin{ytableau} \W \bullet & \W & \W & \W \end{ytableau}
\vee
\begin{ytableau} \W & \W & \W & \W \bullet \end{ytableau}
\end{eqnarray}
For each epistemic state $a\vee b$, we assume that the probability that the 
actual ontic state is $a$ or $b$ is $\frac{1}{2}$ each.

The subset of ontic states $\{a,b\}$ that appear in the epistemic state $a\vee b$ is called
the \textit{ontic support} of the epistemic state.
When the ontic supports of two epistemic states are disjoint, then the two epistemic states are said to be \textit{disjoint}.
$1\vee 2$ and $3\vee 4$, $1\vee 3$ and $2\vee 4$, $2\vee 3$ and $1\vee 4$ are disjoint pairs
of epistemic states.
We will also call one member of these disjoint pairs the \textit{disjoint} of the other.
%The union of ontic supports of two disjoint epistemic states will be the entire
%ontic state space:
%
%\begin{equation}
%1\vee 2\vee 3\vee 4 \;=\; \begin{ytableau} \G & \G & \G & \G \end{ytableau}
%\;.
%\end{equation}
%
%This diagram represents a state of no knowledge about the system.

%%%%%%%%%%%%%%%%%%%%%%%%%%%%%%%%%%%%%%%%%%%%%%%%%%%%%%%%%%%%%%%%%%%%%%%%%%%%%%%%%%%%%%%%%%%%%%%%%%
\subsection{Measurements, Observables, \& Eigenstates}

The only allowed \textit{measurements} in Spekkens' model are yes-no inquiries on whether one of the
$a\vee b$ statements is true or not. 
For instance, if the system is in ontic state 1, the result of a measurement of $1\vee 2$ would result in ``yes,''
while a measurement of $3\vee 4$ would result in ``no.''
It is clear that an ``yes'' answer for ``$1\vee 2$?'' is equivalent to a ``no'' answer for ``$3\vee 4$?'', yielding the same information on whether the system is in state $1\vee 2$ or $3\vee 4$.
Each measurement provides information on whether the system is in one epistemic state or
its disjoint.
Therefore, \textit{observables} can be defined as pairs of disjoint epistemic states, which are
%For the elementary system these would be the three disjoint pairs
$X=\{ 1\vee 3, 2\vee 4 \}$, $Y=\{ 2\vee 3, 1\vee 4 \}$, and $Z=\{ 1\vee 2, 3\vee 4 \}$.

The three observables can be represented graphically as
\begin{eqnarray}
X & \;:\quad & 
\begin{ytableau} \W + & \W - & \W + & \W - \end{ytableau} 
%\begin{ytableau} \W \mathrm{I} & \W \mathrm{II} & \W \mathrm{I} & \W \mathrm{II} \end{ytableau} 
\cr
Y & \;:\quad &
\begin{ytableau} \W - & \W + & \W + & \W - \end{ytableau} 
%\begin{ytableau} \W \mathrm{II} & \W \mathrm{I} & \W \mathrm{I} & \W \mathrm{II} \end{ytableau} 
\cr
Z & \;:\quad &
\begin{ytableau} \W + & \W + & \W - & \W - \end{ytableau} 
%\begin{ytableau} \W \mathrm{I} & \W \mathrm{I} & \W \mathrm{II} & \W \mathrm{II} \end{ytableau} 
\end{eqnarray}
where we denote the two possible outcomes of a measurement of each observable 
for a given underlying ontic state with $+1$ and $-1$.
%I and II.
%Another notation more analogous to spin-systems that Spekkens' model attempts to model may be
%
%\begin{eqnarray}
%\begin{ytableau} \W + & \W - & \W + & \W - \end{ytableau} 
%& \;:\; & \mbox{Observable $X$} \cr
%\begin{ytableau} \W - & \W + & \W + & \W - \end{ytableau} 
%& \;:\; & \mbox{Observable $Y$} \cr
%\begin{ytableau} \W + & \W + & \W - & \W - \end{ytableau} 
%& \;:\; & \mbox{Observable $Z$} 
%\end{eqnarray}
%
The measurement of $Z$, for instance, on the epistemic state $1\vee 2$ always yields $+1$, while on the epistemic state $3\vee 4$ always yields $-1$.
On the other hand,
the measurement of $X$ on the epistemic state $1\vee 2$, which is in ontic state 1 with probability $\frac{1}{2}$ and in ontic state 2 with probability $\frac{1}{2}$, will yield $\pm 1$ with probability $\frac{1}{2}$ each.

Let us call the epistemic states comprising each observable the \textit{eigenstates} of the observable.
For instance, $1\vee 3$ and $2\vee 4$ are the eigenstates of $X$.
A measurement of $X$ on $1\vee 3$ will always yield $+1$ while a measurement of $X$ on $2\vee 4$ will always yield $-1$.

%%%%%%%%%%%%%%%%%%%%%%%%%%%%%%%%%%%%%%%%%%%%%%%%%%%%%%%%%%%%%%%%%%%%%%%%%%%%%%%%%%%%%%%%%%%%%%%%%%
\subsection{Ontic State Update/Epistemic State Collapse}

The knowledge balance principle demands that
repeated measurements cannot be allowed to narrow down the possible underlying ontic state.
The ontic state must therefore be perturbed by each measurement and update accordingly.
For instance,
a $+1$ result on the measurement of $X$ tells us that the system is in ontic state 1 or 3 with $\frac{1}{2}$ probability each.  If the actual ontic state were 1, and did not update as a result of the measurement of $X$, a followup measurement of $Z$ would yield $+1$ with probability 1 and allow us to deduce what the ontic state was.
Therefore, the measurement of $X$ must knock the ontic state 1 into ontic state 3 with $\frac{1}{2}$ probability so that while another measurement of $X$ will yield $+1$ as before, a measurement of $Z$ (or $Y$) will yield $\pm 1$ with probability $\frac{1}{2}$ each.

A measurement will also update the epistemic state.
A measurement of $X$ yielding $+1$ will update the epistemic state to $1\vee 3$ while a $-1$ outcome
will update it to $2\vee 4$.  
In other words, a measurement of an observable will \textit{collapse} the epistemic state to
one of the observable's eigenstates.

%%%%%%%%%%%%%%%%%%%%%%%%%%%%%%%%%%%%%%%%%%%%%%%%%%%%%%%%%%%%%%%%%%%%%%%%%%%%%%%%%%%%%%%%%%%%%%%%%%
\subsection{Correspondence to Qubit/Spin States}

Spekkens' six epistemic states introduced above correspond nicely to qubit/spin states as 
\begin{eqnarray}
\{ 1\vee 2, 3\vee 4 \} & \quad\Leftrightarrow\quad & \{ \kz , \ko \}\;,\cr
\{ 1\vee 3, 2\vee 4 \} & \quad\Leftrightarrow\quad & \{ \kp , \km \}\;,\cr
\{ 2\vee 3, 1\vee 4 \} & \quad\Leftrightarrow\quad & \{ \kpi, \kmi \}\;,
\end{eqnarray}
where $\kz$ (spin up) and $\ko$ (spin down) are the computational basis states of the qubit
(eigenstates of $\hat{Z}=\sigma_z$) while
\begin{equation}
\kpm  \,=\, \dfrac{1}{\sqrt{2}}\Bigl(\kz \pm    \ko \Bigr)\;,\qquad
\kpmi \,=\, \dfrac{1}{\sqrt{2}}\Bigl(\kz \pm i\,\ko \Bigr)\;.
\label{Qsuperpositions}
\end{equation}
%
%\begin{equation}
%\begin{array}{ll}
%\kp  & = \; \dfrac{1}{\sqrt{2}}\Bigl( \kz + \ko \Bigr) \;,\cr
%\km  & = \; \dfrac{1}{\sqrt{2}}\Bigl( \kz - \ko \Bigr) \;,\cr
%\kpi & = \; \dfrac{1}{\sqrt{2}}\Bigl( \kz + i\ko \Bigr) \;,\cr
%\kmi & = \; \dfrac{1}{\sqrt{2}}\Bigl( \kz - i\ko \Bigr) \;.
%\end{array}
%\label{Qsuperpositions}
%\end{equation}
%
The states $\kpm$ and $\kpmi$ are respectively the eigenstates of
$\hat{X}=\sigma_x$ and $\hat{Y}=\sigma_y$ with eigenvalues $\pm 1$.
%In Figure~\ref{bloch} we show the position of these states on the Bloch sphere.
%
Given this correspondence,
it is straightforward to show that the measurement outcomes of $X$, $Y$, $Z$ on Spekkens' six epistemic states correspond exactly to the measurement outcomes of $\hat{X}$, $\hat{Y}$, $\hat{Z}$ on the six qubit/spin states.
In other words, Spekkens' toy model in its most elementary form succeeds in
reproducing the predictions of a spin-$\frac{1}{2}$ quantum mechanical system albeit 
restricted to the six states listed here.

%%%%%%%%%%%%%%%%%%%%%%%%%%%%%%%%%%%%%%%%%%%%%%%%%%%%%%%%%%%%%%%%%%%%%%%%%%%%%%%%%%%%%%%%%%%%%%%%%%
\subsection{Entanglement of two ``Qubits/Spins''}

\begin{figure}[t]
\begin{center}
\includegraphics[height=3cm]{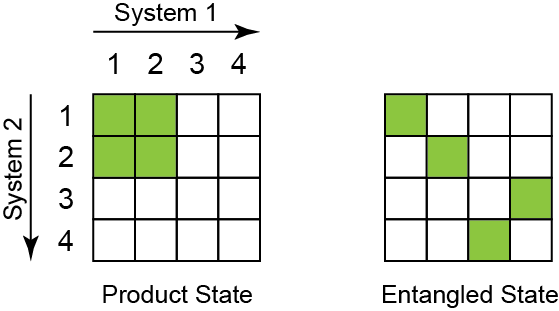}
\caption{Examples of product and entangled states in Spekkens' model.
}
\label{product_and_entangled_states}
\end{center}
\end{figure}

A pair of elementary systems, each with $2^2=4$ ontic states, will have $2^4=16$ ontic states.
A complete specification of the ontic state will require 4 bits of information, whereas
the knowledge balance principle demands that only 2 bits of information may be known about the
system at any time.  
Furthermore, the knowledge balance principle continues to apply to each of the 2-bit subsystems.
This restricts the epistemic states of the elementary-system pair to the types depicted graphically
in Figure~\ref{product_and_entangled_states}:
there are $6^2=36$ product states, and $4!=24$ entangled states.
The entangled states must have one shaded box in each row and each column.
Each measurement of either system 1 or system 2 will update the ontic and epistemic states
as in the one-system case. 
Any measurement on an entangled state will collapse it to a product state.

%%%%%%%%%%%%%%%%%%%%%%%%%%%%%%%%%%%%%%%%%%%%%%%%%%%%%%%%%%%%%%%%%%%%%%%%%%%%%%%%%%%%%%%%%%%%%%%%%%
%%%%%%%%%%%%%%%%%%%%%%%%%%%%%%%%%%%%%%%%%%%%%%%%%%%%%%%%%%%%%%%%%%%%%%%%%%%%%%%%%%%%%%%%%%%%%%%%%%

%%%%%%%%%%%%%%%%%%%%%%%%%%%%%%%%%%%%%%%%%%%%%%%%%%%%%%%%%%%%%%%%%%%%%%%%%%%%%%%%%%%%%%%%%%%%%%%%%%
\subsection{Linearity}

In Ref.~\cite{Spekkens:2007},
Spekkens defines four types of ``coherent superpositions'' as
\begin{equation}
\begin{array}{l}
(a\vee b) +_1     (c\vee d) \;=\; a \vee c\;,\cr
(a\vee b) +_2     (c\vee d) \;=\; b \vee d\;,\cr
(a\vee b) +_3     (c\vee d) \;=\; b \vee c\;,\cr
(a\vee b) +_4     (c\vee d) \;=\; a \vee d\;.
\end{array}
\label{FOILsums0}
\end{equation}
If applied to $(1\vee 2)$ and $(3\vee 4)$, we have
\begin{equation}
\begin{array}{l}
(1\vee 2) +_1     (3\vee 4) \;=\; 1 \vee 3\;,\cr
(1\vee 2) +_2     (3\vee 4) \;=\; 2 \vee 4\;,\cr
(1\vee 2) +_3     (3\vee 4) \;=\; 2 \vee 3\;,\cr
(1\vee 2) +_4     (3\vee 4) \;=\; 1 \vee 4\;,
\end{array}
\label{FOILsums1}
\end{equation}
in exact parallel to Eq.~\eqref{Qsuperpositions}.
Thus, these four different sums are analogous to addition but with
different phases multiplying the second state of the sum.

However, note that these sums are only defined for pairs of disjoint
epistemic states. The sum of $1\vee 2$ and $1\vee 3$, for instance, is not defined.
Moreover, the sums are dependent on the ordering of the ontic-state labels.  
One could try to identify $a\vee b$ and $b\vee a$ with the same qubit/spin 
state but with different phases, but it can be shown that such
an identification would not work.

Since one cannot define arbitrary sums of the six
qubit/spin states $\ket{0}$, $\ket{1}$, $\kpm$, and $\kpmi$ in canonical QM either
without going outside of this set,
one could argue that the inability to define sums of arbitrary pairs of epistemic states
in Spekkens' model is not really a problem.
Nevertheless, the lack of linearity suggests that an important aspect
of QM has not been completely captured within the model.

%%%%%%%%%%%%%%%%%%%%%%%%%%%%%%%%%%%%%%%%%%%%%%%%%%%%%%%%%%%%%%%%%%%%%%%%%%%%%%%%%%%%%%%%%%%%%%%%%%
\section{Finite Field Quantum Mechanics}

The missing linearity of Spekkens' toy model can be supplied by mapping it to a
QM defined over the finite field $\mathbb{F}_5$.
There are multiple ways to define QM on finite fields as discussed in
Refs.~\cite{GQM:2013,GQM-Spin:2013,Chang:2013,Chang:2014,QFun:2014}.
The formalism presented here is closest to Refs.~\cite{GQM:2013,GQM-Spin:2013}.

The finite field $\mathbb{F}_5 = \{\0,\pm\1,\pm\2\}$ consists of five elements
with addition and multiplication defined modulo 5. 
We underline elements of $\mathbb{F}_5$ to distinguish them from elements of $\mathbb{Z}$.
Note that $\2^2 = (-\2)^2 = -\1$, that is:
\begin{equation}
\pm\2 \;=\; \pm\sqrt{-\1}\;.
\end{equation}
%
%Note also that $\2(-\2) = (-\2)\2 = \1$, that is:
%
%\begin{equation}
%\pm\2 \;=\; \mp\1/\2\;.
%\end{equation}
%
In analogy to the qubit/spin states in canonical QM, which are vectors 
in the complex projective space $CP^1 = P^1(\mathbb{C})$ (the Bloch sphere), 
the ``qubit/spin'' states in $\mathbb{F}_5$ QM are vectors in the $\mathbb{F}_5$ projective space $P^1(\mathbb{F}_5)$. 
This is basically the 2D vector space $\mathbb{F}_5^2$ with vectors differing by overall non-zero multiplicative constants
identified as representing the same state.
Consequently, in $P^1(\mathbb{F}_5)$ 
there are $(5^2-1)/4=6$ inequivalent states and these can be taken to be
%
%Thus, of the $5^2-1=24$ non-zero vectors in $\mathbb{F}_5^2$, every four of them are equivalent, and the $24/4=6$ inequivalent ones can be taken to be:\footnote{%The notation we use here is different from the one we used in Ref.~\cite{Chang:2012gg}.}
\begin{equation}
\ket{a} = \left[\begin{array}{r} \1 \\  \0 \end{array}\right],\;
\ket{b} = \left[\begin{array}{r} \0 \\  \1 \end{array}\right],\;
\ket{c} = \left[\begin{array}{r} \1 \\  \1 \end{array}\right],\;
\ket{d} = \left[\begin{array}{r} \1 \\ -\1 \end{array}\right],\;
\ket{e} = \left[\begin{array}{r} \1 \\  \2 \end{array}\right],\; 
\ket{f} = \left[\begin{array}{r} \1 \\ -\2 \end{array}\right].    
\end{equation}
The inequivalent dual-vectors can be taken to be
\begin{eqnarray}
\bra{a} & = & \bigl[\;\1\;\;\0 \;\bigr]\;,\quad
\bra{b} \;=\; \bigl[\;\0\;\;\1 \;\bigr]\;,\quad
\bra{c} \;=\; \bigl[\;-\2\;-\!\2 \;\bigr]\;,\cr
\bra{d} & = & \bigl[\;-\2\;\;\2 \;\bigr]\;,\quad
\bra{e} \;=\; \bigl[\;-\2\;-\!\1 \;\bigr]\;,\quad
\bra{f} \;=\; \bigl[\;-\2\;\;\1 \;\bigr]\;.
\end{eqnarray}
%\begin{eqnarray}
%\bra{\bar{a}} & = & \bigl[\;\0\;-\!\1           \;\bigr]\;,\cr
%\bra{\bar{b}} & = & \bigl[\;\1\;\;\phantom{-}\0 \;\bigr]\;,\cr
%\bra{\bar{c}} & = & \bigl[\;\1\;-\!\2           \;\bigr]\;,\cr
%\bra{\bar{d}} & = & \bigl[\;\1\;\;\phantom{-}\1 \;\bigr]\;,\cr
%\bra{\bar{e}} & = & \bigl[\;\1\;\;\phantom{-}\2 \;\bigr]\;,\cr
%\bra{\bar{f}} & = & \bigl[\;\1\;-\!\1           \;\bigr]\;.
%\end{eqnarray}
%
The actions of these dual-vectors on the vectors are:
\smallskip
\begin{center}
\begin{tabular}{|c||cc|cc|cc|}
\hline
& $\quad\ket{a}\quad$ & $\quad\ket{b}\quad$ & $\quad\ket{c}\quad$ & $\quad\ket{d}\quad$ & $\quad\ket{e}\quad$ & $\quad\ket{f}\quad$ \\
\hline
$\;\;\bra{a}\;\;$ & $\phantom{-}\1$ & $\phantom{-}\0$ & $\phantom{-}\1$ & $\phantom{-}\1$ & $\phantom{-}\1$ & $\phantom{-}\1$ \\
$\bra{b}$         & $\phantom{-}\0$ & $\phantom{-}\1$ & $\phantom{-}\1$ &           $-\1$ & $\phantom{-}\2$ &           $-\2$ \\
\hline
$\bra{c}$         &           $-\2$ &           $-\2$ & $\phantom{-}\1$ & $\phantom{-}\0$ &           $-\1$ & $\phantom{-}\2$ \\
$\bra{d}$         &           $-\2$ & $\phantom{-}\2$ & $\phantom{-}\0$ & $\phantom{-}\1$ & $\phantom{-}\2$ &           $-\1$ \\
\hline
$\bra{e}$         &           $-\2$ &           $-\1$ & $\phantom{-}\2$ &           $-\1$ & $\phantom{-}\1$ & $\phantom{-}\0$ \\
$\bra{f}$         &           $-\2$ & $\phantom{-}\1$ &           $-\1$ & $\phantom{-}\2$ & $\phantom{-}\0$ & $\phantom{-}\1$ \\
\hline
\end{tabular}
\end{center}
Observables are defined by pairs of dual-vectors:
\begin{equation}
\underline{X} = \{ \bra{c}, \bra{d}\} \;,\qquad
\underline{Y} = \{ \bra{e}, \bra{f}\} \;,\qquad
\underline{Z} = \{ \bra{a}, \bra{b}\} \;,
\end{equation}
where the first (second) dual-vector of the pair corresponds to the outcome $+1$ ($-1$).
The probability of obtaining outcome $+1$ ($-1$) when observable $O=\{\bra{x},\bra{y}\}$ is measured on state $\ket{\psi}$ 
is defined as
\begin{equation}
P(+1,O,\psi) \,=\, \dfrac{|\braket{x}{\psi}|^2}{|\braket{x}{\psi}|^2+|\braket{y}{\psi}|^2}\;,\qquad
P(-1,O,\psi) \,=\, \dfrac{|\braket{y}{\psi}|^2}{|\braket{x}{\psi}|^2+|\braket{y}{\psi}|^2}\;,\qquad
\end{equation}
where the absolute value of $\underline{k}\in\mathbb{F}_5$ is defined to be $|\underline{k}|=0$ if $\underline{k}=\0$,
and $|\underline{k}|=1$ otherwise.
It is straightforward to demonstrate that the predictions of $\mathbb{F}_5$ QM agrees with those of canonical QM and Spekkens' toy model 
via the correspondence
\begin{equation}
\begin{array}{lll}
\{ 1\vee 2, 3\vee 4 \} & \Leftrightarrow\;\; \{ \kz , \ko \}  & \Leftrightarrow\;\; \{ \ket{a},\ket{b} \} \;,\\
\{ 1\vee 3, 2\vee 4 \} & \Leftrightarrow\;\; \{ \kp , \km \}  & \Leftrightarrow\;\; \{ \ket{c},\ket{d} \} \;,\\
\{ 2\vee 3, 1\vee 4 \} & \Leftrightarrow\;\; \{ \kpi, \kmi \} & \Leftrightarrow\;\; \{ \ket{e},\ket{f} \} \;.
\end{array}
\label{SpekkensF5correspondence}
\end{equation}
However, in $\mathbb{F}_5$ QM, the sums of arbitrary pairs of states are well defined.
In addition to
\begin{eqnarray}
\ket{c} & = & \ket{a} + \ket{b}\;,\cr
\ket{d} & = & \ket{a} - \ket{b}\;,\cr
\ket{e} & = & \ket{a} + \2\,\ket{b} \;=\; \ket{a} + \sqrt{-\1}\,\ket{b}\;,\cr
\ket{f} & = & \ket{a} - \2\,\ket{b} \;=\; \ket{a} - \sqrt{-\1}\,\ket{b}\;,
\end{eqnarray}
we also have, for instance:
\begin{eqnarray}
\ket{a} + \ket{c} & = & \2\ket{f}\;,\cr
\ket{a} - \ket{c} & = & -\ket{b}\;,\cr
\ket{a} + \2\,\ket{c} & = & -\2\ket{d}\;,\cr
\ket{a} - \2\,\ket{c} & = & -\ket{e}\;,
\end{eqnarray}
which corresponds to
\begin{eqnarray}
(1\vee 2)+_1(1\vee 3) & = & (1\vee 4)\;,\cr
(1\vee 2)+_2(1\vee 3) & = & (3\vee 4)\;,\cr
(1\vee 2)+_3(1\vee 3) & = & (2\vee 4)\;,\cr
(1\vee 2)+_4(1\vee 3) & = & (2\vee 3)\;.
\end{eqnarray}
In this fashion, one can introduce sums of non-disjoint epistemic states 
into Spekkens' toy model via the correspondence Eq.~\eqref{SpekkensF5correspondence}.

%%%%%%%%%%%%%%%%%%%%%%%%%%%%%%%%%%%%%%%%%%%%%%%%%%%%%%%%%%%%%%%%%%%%%%%%%%%%%%%%%%%%%%%%%%%%%%%%%%
%\newpage

\begin{figure}[t]
\begin{center}
\includegraphics[height=4.5cm]{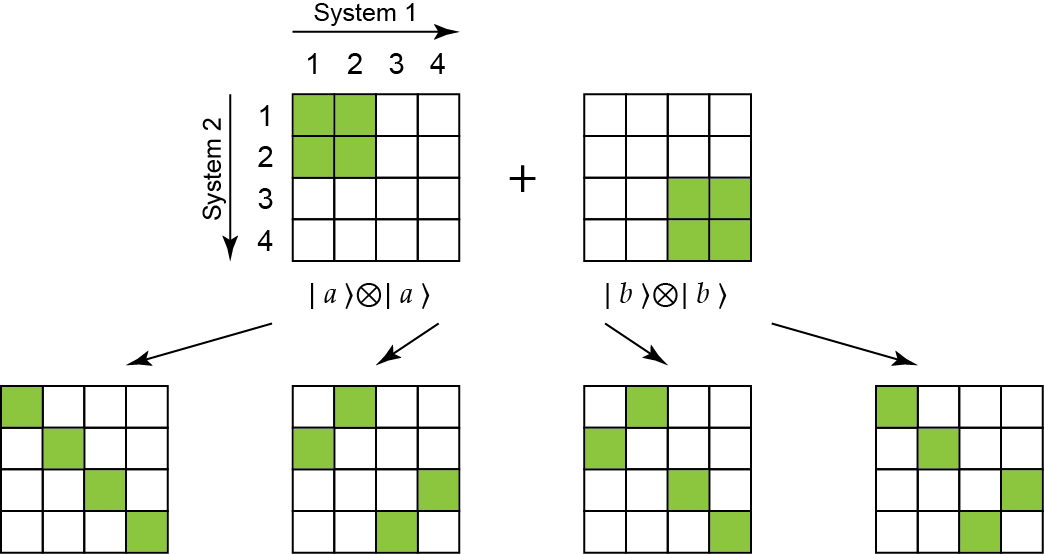}
\caption{Possible ``coherent superpositions'' of disjoint epistemic states.}
\label{DisjointSums}
\end{center}
\end{figure}

\section{Entanglement in $\mathbb{F}_5$ QM and in Spekkens' Toy Model}

Since $\mathbb{F}_5$ QM is equipped with a vector space, the two-qubit/spin system state space
is obtained by tensoring the one-qubit/spin system state space. This gives us $P^3(\mathbb{F}_5)$,
namely $\mathbb{F}_5^4$ with vectors differing by overall non-zero multiplicative constants identified.
The number of inequivalent states in this space is $(5^4-1)/4=156$, of which $6^2=36$ are product states and
$156-36=120$ are entangled.
All 120 entangled states can be expressed as superpositions of two product states
without a common factor.

There is clearly a mismatch in the number of entangled states between $\mathbb{F}_5$ QM and Spekkens' model.
Thus, imposing linearity on Spekkens' model seems to require many more entangled epistemic states to be included 
in the model than were originally allowed. 
So which ones are they?  
And can we prevent the inclusion of such states from violating the knowledge balance principle,
perhaps by modifying the ontic-state update rule upon measurement?

Before pursuing these questions further, let us first identify
which of the 120 $\mathbb{F}_5$ entangled states correspond to the 24 Spekkens' entangled states.
Do all Spekkens' entangled states have an $\mathbb{F}_5$ analog?
Considering how ``coherent superpositions'' were originally defined for the elementary system, one guess would be that
the Spekkens' entangled states would result from sums of disjoint product states as depicted in Figure~\ref{DisjointSums}.
Indeed, for the disjoint pair $\{\ket{a}\otimes\ket{a},\ket{b}\otimes\ket{b}\}$ one finds
\begin{equation}
\begin{array}{lll}
\ket{a}\otimes\ket{a}+\ket{b}\otimes\ket{b} 
& = \; -2(\ket{c}\otimes\ket{c}+\ket{d}\otimes\ket{d}) 
& = \; -2(\ket{e}\otimes\ket{f}+\ket{f}\otimes\ket{e}) 
\;,\\
\ket{a}\otimes\ket{a}-\ket{b}\otimes\ket{b} 
& = \; -2(\ket{c}\otimes\ket{d}+\ket{d}\otimes\ket{c}) 
& = \; -2(\ket{e}\otimes\ket{e}+\ket{f}\otimes\ket{f}) 
\;,\\
\ket{a}\otimes\ket{a}+\2\ket{b}\otimes\ket{b} 
& = \; -2(\ket{c}\otimes\ket{e}+\ket{d}\otimes\ket{f}) 
& = \; -2(\ket{e}\otimes\ket{c}+\ket{f}\otimes\ket{d}) 
\;,\\
\ket{a}\otimes\ket{a}-\2\ket{b}\otimes\ket{b} 
& = \; -2(\ket{c}\otimes\ket{f}+\ket{d}\otimes\ket{e}) 
& = \; -2(\ket{f}\otimes\ket{c}+\ket{e}\otimes\ket{d}) 
\;.
\end{array}
\label{aabbsuper}
\end{equation}

\begin{figure}[t]
\begin{center}
\includegraphics[height=4.2cm]{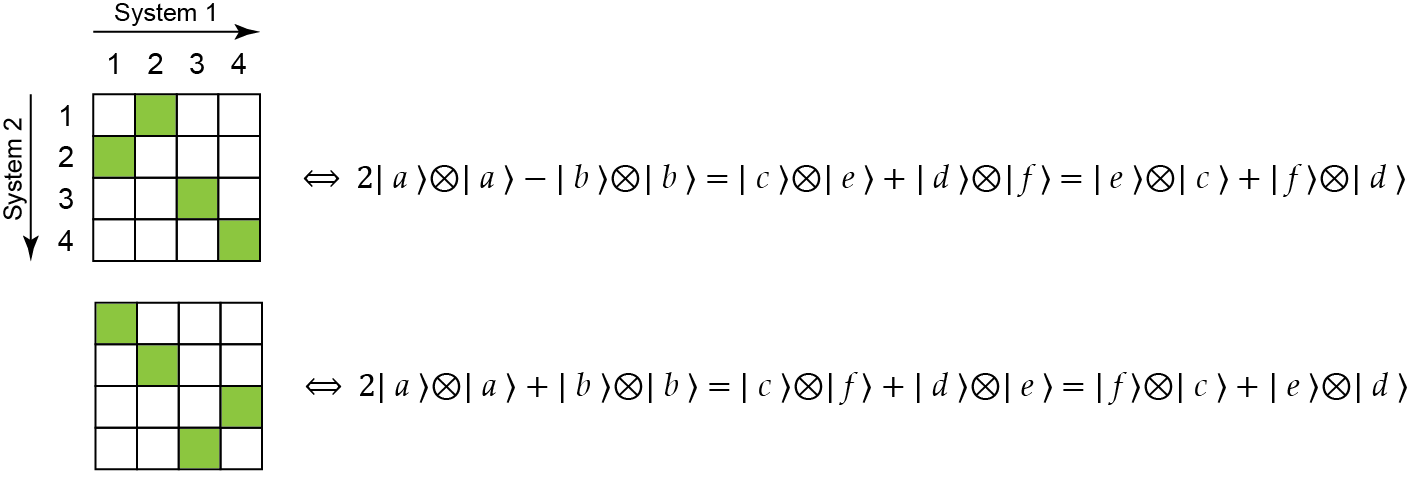}
\caption{Spekkens' entangled states and the corresponding $\mathbb{F}_5$ entangled states.}
\label{aapm2bb}
\end{center}
\end{figure}

\begin{figure}[b]
\begin{center}
\includegraphics[width=12cm]{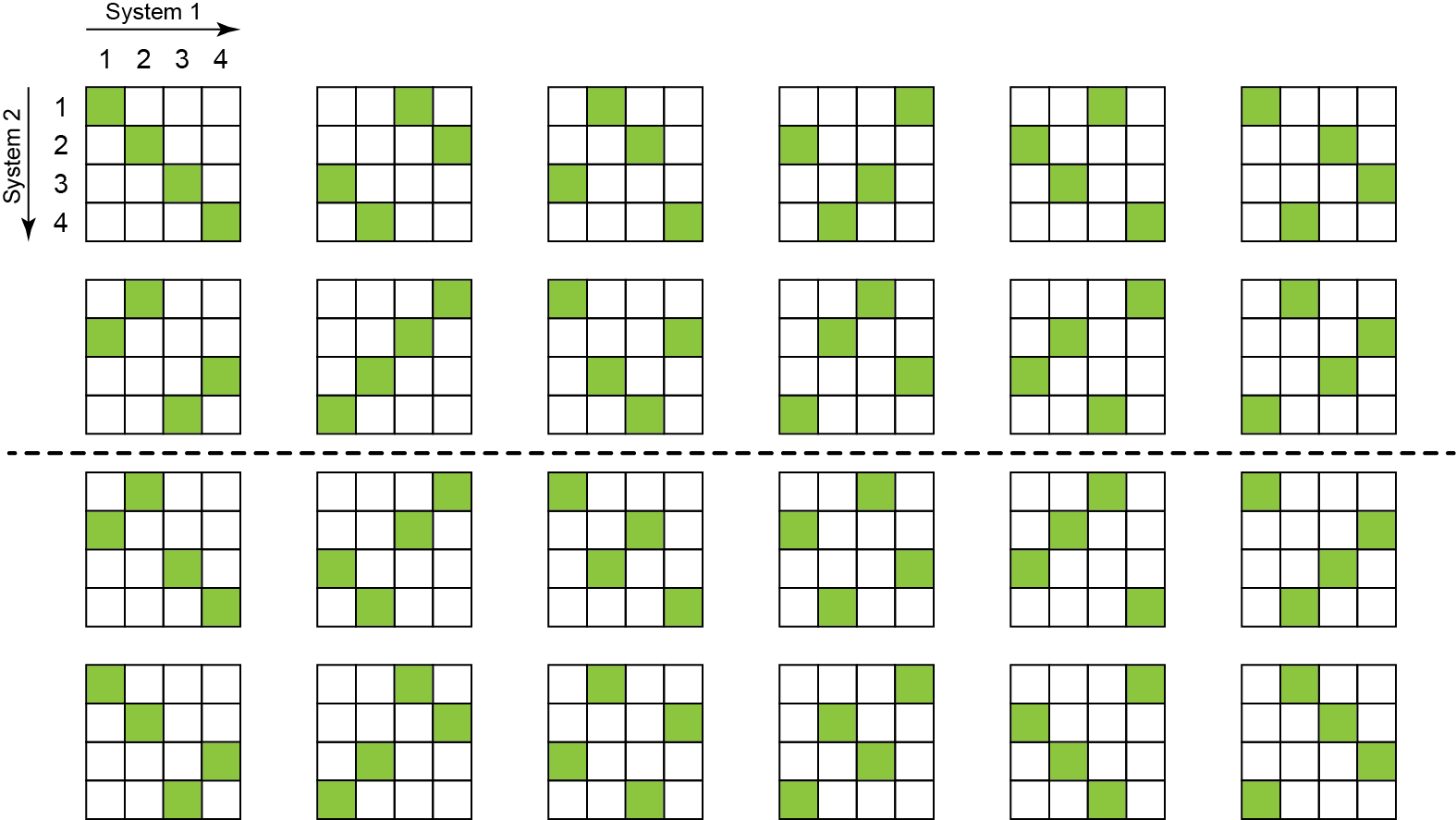}
\caption{The $4!=24$ entangled states of the Spekkens' model.
The 12 states above the dashed line do not have $\mathbb{F}_5$ analogs.}
\label{entangled_states}
\end{center}
\end{figure}

\noindent
Comparing the result of various measurements on these states with those of Spekkens' states, 
one finds that $\ket{a}\otimes\ket{a}\pm\2\,\ket{b}\otimes\ket{b}$ have Spekkens' analogs, cf. Figure~\ref{aapm2bb}, 
but $\ket{a}\otimes\ket{a}\pm \ket{b}\otimes\ket{b}$ do not.
For instance, the Spekkens' state
\begin{equation}
\begin{ytableau}
\G & \W & \W  & \W \\
\W & \G & \W  & \W \\
\W & \W & \G  & \W \\
\W & \W & \W  & \G \\
\end{ytableau}
\end{equation}
will collapse to the product state $\ket{a}\otimes\ket{a}$ or $\ket{b}\otimes\ket{b}$ if $Z$ is measured
on system 1 or 2, $\ket{c}\otimes\ket{c}$ or $\ket{d}\otimes\ket{d}$ if $X$ is measured on system 1 or 2,
and $\ket{e}\otimes\ket{e}$ or $\ket{f}\otimes\ket{f}$ if $Y$ is measured on system 1 or 2.
However, as we have demonstrated in Eq.~\eqref{aabbsuper}, no superposition of 
$\ket{a}\otimes\ket{a}$ and $\ket{b}\otimes\ket{b}$ is also a superposition of 
$\ket{c}\otimes\ket{c}$ and $\ket{d}\otimes\ket{d}$,
and a superposition of $\ket{e}\otimes\ket{e}$ and $\ket{f}\otimes\ket{f}$ simultaneously.
Indeed, of the $4!=24$ entangled states of Spekkens' model, 
12 do not have $\mathbb{F}_5$ analogs that give the same predictions, cf. Figure~\ref{entangled_states}.

The reason for this mismatch can be traced to the fact that the correspondence of
Eq.~\eqref{SpekkensF5correspondence} does not respect the symmetries of the respective models.
Spekken's toy model is invariant under any relabelling of the four ontic states.
In other words, it has an $S_4$ symmetry.
On the other hand, the qubit/spin states of both canonical QM and $\mathbb{F}_5$ QM transform 
under 2D projective representations of $SO(3)$.
Though $S_4$ can be embedded into $SO(3)$ \cite{Klein}, 
Eq.~\eqref{SpekkensF5correspondence} demands that
all the even permutations of $S_4$ (the $A_4$ subgroup) 
be mapped to rotations, but all the odd permutations be mapped to reflections, which would involve
a complex conjugation.
In fact, this mismatch was already evident in Ref.~\cite{Spekkens:2007} 
where it was noted that the coherent superpositions of Eq.~\eqref{FOILsums0}
do not produce the expected results for sums of $1\vee 3$ and $2\vee 4$.

Thus, we have identified a rather serious problem 
if one wishes to utilize Spekkens' model as a model of qubits/spin.
While the linearity requirement demanding more entangled epistemic states to be added to the mix is interesting in itself,
the symmetry mismatch suggests we could be comparing apples and oranges.
One possible way out may be to embed the $S_4$ permutations into 4D projective representations of $SO(3)$ 
in canonical and $\mathbb{F}_5$ QM.
This will allow the odd permutations to be also represented as rotations, but it will also 
double the number of states available leading to a different set of complications.
These, and other questions will be addressed in more detail in a forthcoming paper.

%%%%%%%%%%%%%%%%%%%%%%%%%%%%%%%%%%%%%%%%%%%%%%%%%%%%%%%%%%%%%%%%%%%%%%%%%%%%%%%%%%%%%%%%%%%%%%%%%%
%\begin{figure}[t]
%\includegraphics[width=6cm]{BlochSphere.pdf}
%\caption{The Bloch Sphere and the six qubit (spin) states that correspond to Spekkens' six
%epistemic states.
%}
%\label{bloch}
%\end{figure}
%%%%%%%%%%%%%%%%%%%%%%%%%%%%%%%%%%%%%%%%%%%%%%%%%%%%%%%%%%%%%%%%%%%%%%%%%%%%%%%%%%%%%%%%%%%%%%%%%%

%%%%%%%%%%%%%%%%%%%%%%%%%%%%%%%%%%%%%%%%%%%%%%%%%%%%%%%%%%%%%%%%%%%%%%%%%%%%%%%%%%%%%%%%%%%%%%%%%%%%
\ack
Takeuchi would like to thank the Institute of Advanced Studies at Nanyang Technological University, Singapore, 
for their hospitality during the QM90 Conference (23-26 January 2017), 
and Sixia Yu for bringing Spekkens' model to his attention.  
We would also like to thank Chen Sun and Matthew Henriques for their respective contributions to this project.

%%%%%%%%%%%%%%%%%%%%%%%%%%%%%%%%%%%%%%%%%%%%%%%%%%%%%%%%%%%%%%%%%%%%%%%%%%%%%%%%%%%%%%%%%%%%%%%%%%%%

\end{document}